# Technical Report: Benchmarking tunnel and encryption methodologies in cloud environments

PRAVEIN GOVINDAN KANNAN, IBM Research - India
BRENT SALISBURY, Red Hat
PALANIVEL KODESWARAN, IBM Research - India
SAYANDEEP SEN, IBM Research - India

The recent past has seen the adoption of multi-cloud deployments by enterprises due to availability, features and regulatory requirements. A typical deployment involves parts of an application/workloads running inside a private cloud with the other parts spread across multiple on-prem/public clouds. Typical cluster-to-cluster networking in such deployments involve the establishment of site-to-site encrypted tunnels to connect the workloads.

In this report, we benchmark the performance of various tunneling and encryption technologies to provide directions on their use in multi-cloud deployments. Based on the various experiments conducted on three different testbeds, we present quantifiable data which can be leveraged by operators and cloud providers tasked with design and development decisions of multi-cloud network connectivity and orchestration.

## 1 INTRODUCTION

In the past decade, we have seen increasing adoption of the cloud by several businesses. Recently, we are observing yet another wave of transition towards edge-computing and single-cloud to multi-cloud, i.e., businesses preferring to use multiple cloud providers[3] for their workloads due to availability, features, and regulatory requirements. This stresses the importance of transitioning private traffic between multiple clusters provided by different cloud providers. The de-facto way of doing this is to employ a tunneling mechanism between each individual pair of clusters (as done by Submariner [14], Aviatrix [2], etc) that does not require configuration changes to the underlying network and provides the applications a seamless way to connect using a private (RFC 1918) IP addressing scheme. Figure 1 illustrates how an application could be deployed in multiple cloud clusters which have pre-established tunnels based on the application's connectivity requirements. In certain scenarios involving privacy and security concerns (HIPAA, PCI DSS [6]), there are requirements to additionally encrypt the traffic between the clusters. Hence, tunneling and encryption are the main primitives that enable multi-cloud networking, and it is essential to benchmark these primitives over different deployment scenarios to provide directions on their design/deployment. While there have been past efforts [12, 17] to benchmark tunnels and encryption, they are done on localized setup, and do not present the variation of performance w.r.t their deployment scenarios.

In this report, we systematically evaluate popular overlay tunneling and encryption mechanisms over different setups (inter-DC VMs over WAN, intra-DC VMs within a data-center and an on-prem intra-rack bare-metal setup) to provide insights on their performance (throughput & latency) and the overhead (CPU utilization) incurred. We performed the experiments using standard open-source performance benchmarking tools. We hope this report provides quantifiable data to operators and cloud providers tasked with design and development decisions of multi-cloud network connectivity and orchestration.

Our objective from this benchmarking exercise is to measure the throughput degradation due to various tunnel and encryption techniques using off-the-shelf configurations on various experimental setups. All of the tools used in this project were open-source, along with running the experiments on standard Linux. The tunnel setup and teardown for each scenario were all automated via scripts that will be open-sourced once documented.

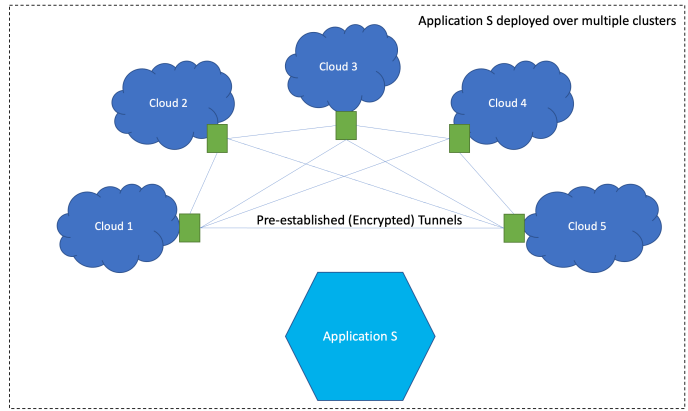

Fig. 1. An Application S deployed over multiple cloud environments which are connected using pre-established tunnels with encryption

This report is organized as follows. We present our testbed setup with the tools used for performing measurements in Section 2. We present our measurements results in Section 3.1, 3.2 and 3.3. Finally, we present discussion and conclusion in Section 4.

## 2 TESTBED SETUP

We perform measurements on three widely different testbeds:

(1) **Inter-DC Setup**: A pair of VMs (OS: Ubuntu 20.04.4 LTS; Kernel version: 5.4.0-91-generic) which are connected over WAN (i.e., a VM located in London and the other in Chennai) and equipped with 8-core Intel(R) Xeon(R) Platinum 8260 CPU @ 2.40GHz unaware of the network bandwidth provisioned to the VM.

(2) **Intra-DC Setup**: A pair of VMs (OS: Ubuntu 20.04.4 LTS; Kernel version: 5.4.0-1047-kvm) within the same data-center (i.e., both VMs located in the same DC cluster in Sydney) with 16-core Intel Xeon @2.5 GHz with network bandwidth of 16 Gbps provisioned automatically.

(3) **Intra-rack Setup**: A pair of bare-metal servers (OS: Ubuntu 20.04.3 LTS; kernel version: 5.4.0-89-generic) connected via a ToR (Top-of-the-Rack switch), equipped with 80-core Intel(R) Xeon(R) Gold 5218R CPU @ 2.10 GHz and connected to each other using 40G QSFP+ DAC cables via Mellanox Connect-X5 NICs [7].

We made use of following benchmarking tools for throughput and performance:

(1) Netperf[10] to benchmark throughput and latency of a single TCP flow across the nodes.
(2) iperf3[4] to benchmark effect on throughput due to multiple parallel flows.
(3) mpstat[8] to measure the overall CPU utilization.

We benchmark the performance of the following commonly used tunnel/encryption technologies:

(1) Native: The baseline native NIC interface without adding any tunnels.
(2) Geneve: A Geneve interface setup using netlink helpers [9] available in golang.
(3) VXLAN: A VXLAN interface setup using netlink helpers [9].
(4) OVS+Geneve: A Geneve interface setup by creating a network namespace within an OVS (OpenVSwitch) [11] bridge.
(5) OVS+VXLAN: A VXLAN interface setup by creating a network namespace within an OVS bridge.
(6) Wireguard: An encryption interface setup using wireguard [15].

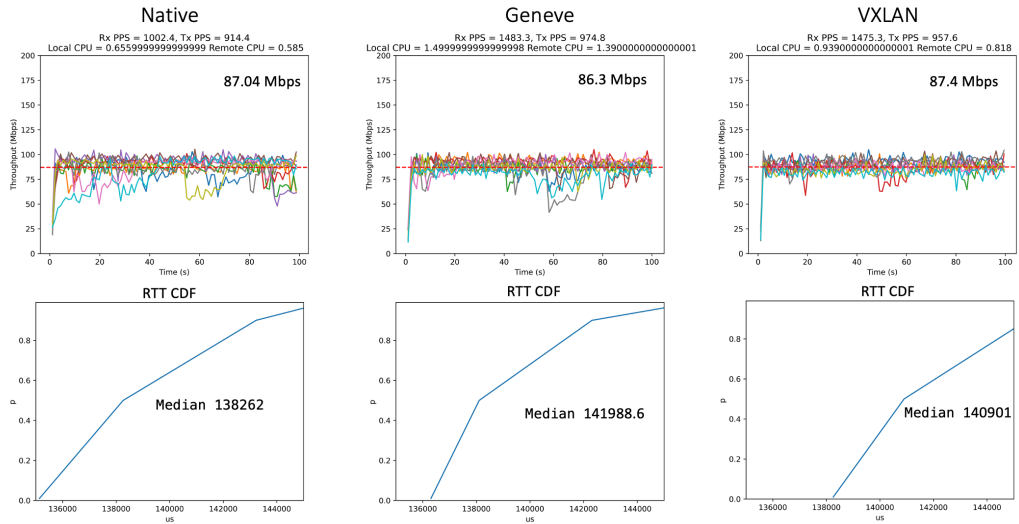

Fig. 2. Throughput series and RTT CDF of the Native, Geneve and VXLAN interfaces for a single stream netperf in inter-DC setup

(7) IPSec: Tunnel-mode IPSec setup using libreswan [5].

The tunnel setup, teardown and the benchmarking measurements were automated using go and shell scripts. The scripts create an additional interface for the tunnels/encryption. The objective of this evaluation is to plug-in the various tunneling/encryption mechanisms to the different setups without any specific customization or tuning to the OS/NIC/etc. to observe their native performance.

## 3 EVALUATION

In this section, we present our results and observations from the inter-DC VM setup in Section 3.1, intra-DC VM setup in Section 3.2 and the inra-rack bare-metal setup in Section 3.3.

### 3.1 Inter-DC Measurements

We start the measurements on a pair of inter-DC VMs across continents. The VMs are hosted on IBM's SoftLayer [13] Cloud with one VM hosted in London DC and the other in Chennai DC. Both VMs are equipped with 8-core Intel(R) Xeon(R) Platinum 8260 CPU @ 2.40GHz. We did not have any way to provision any specific network bandwidth to the VM. We conduct ten separate TCP runs in a sequence for each interface separately. We use the following commands to generate throughput series, and RTT:

```
Netperf −H <Destination > −l 100 −D 1 −t TCP_STREAM
    −c −C −− −o local_cpu_util , remote_cpu_util −R 1
```
Listing 1. Command to generate throughput series

```
Netperf −H <Destination > −l 100 −D 1 −t TCP_RR
    −c −C −− −o min_latency , p50_latency , p90_latency ,
    p99_latency , max_latency −R 1
```
Listing 2. Command to get RTT variation

Figure 2, 3 and 4 show the throughput series, and the CDF of RTTs observed for the various interfaces. The different colors in the throughput series graph represents individual 10 runs, and the red dashed line, the average throughput. We observe that there is not much difference or degradation

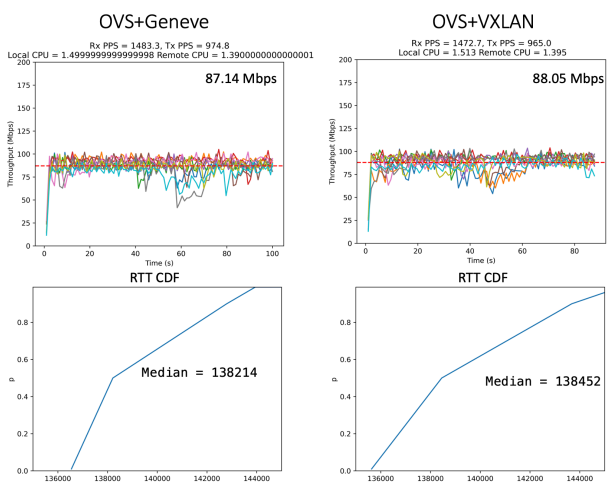

Fig. 3. Throughput series and RTT CDF of the OVS+Geneve and OVS+VXLAN interfaces for a single stream netperf in inter-DC setup

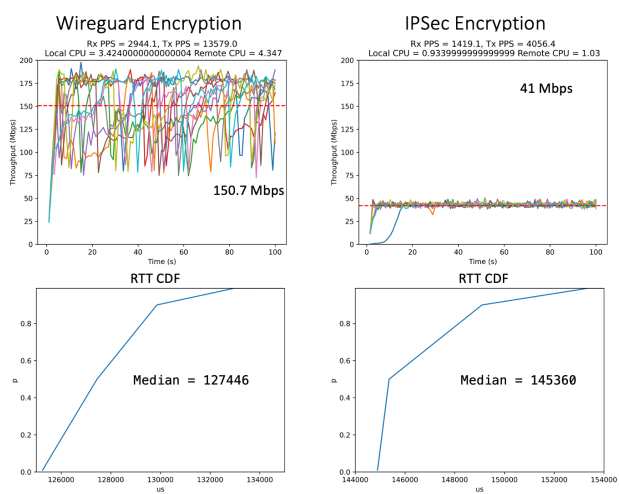

Fig. 4. Throughput series and RTT CDF of the Wireguard and IPSec interfaces for a single stream netperf in inter-DC setup

in throughput of a single TCP stream due to tunnelling interfaces (Geneve, VXLAN, OVS+GENEVE and OVS+VXLAN) compared to the native interface which we observe to be about 87-88 Mbps. Additionally, we do not observe consistent increase in RTTs due to tunnelling since VM overhead and WAN links have larger latency and fluctuations. However, with the encryption interfaces, the behaviour is quite different. While, with Wireguard interface we observe the throughput increase to about 150 Mbps, with IPSec the throughput degrades to about 40 Mbps. Additionally, we observe that the RTT increases by few milliseconds for IPSec, while the RTT decreases with Wireguard interfaces. To understand this behaviour, we plot the average CPU utilization observed by netperf in Figure 5. We can observe that Wireguard utilizes more CPU since it deploys separate threads which run over multiple cores. Due to this, Wireguard manages to drain the RX queue faster. However, IPSec uses fewer CPU since it runs only on a single Core[1]. We observe a marginal increase in average CPU utilization with the Geneve, VXLAN and their OVS counterparts.

---

[1]Verified using htop, where only one core is used.

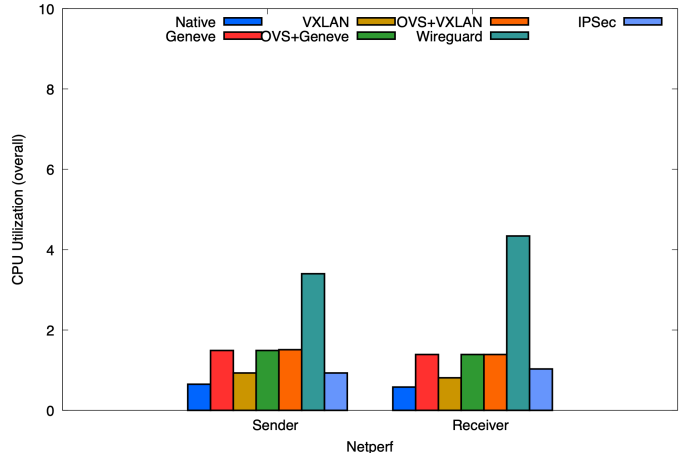

Fig. 5. Average CPU utilization observed by netperf sender/receiver during the measurements

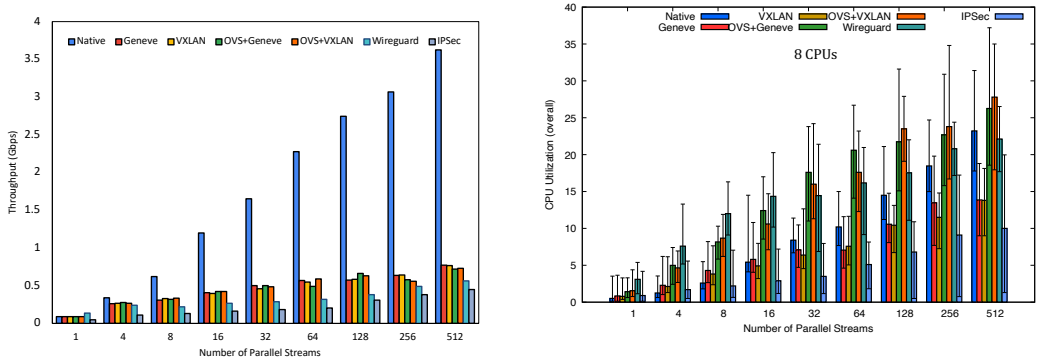

(a) Aggregate throughput observed using iperf3 for multiple parallel TCP streams

(b) Average CPU utilization observed using Mpstat for multiple parallel TCP streams

Fig. 6. Throughput and CPU utilization for multiple parallel TCP streams

Next, we scale the measurement beyond a single TCP Stream to multiple parallel streams. We use iperf3 for this purpose since it supports up to 128 parallel streams. To scale beyond 128, we deploy additional iperf3 processes. For example, for 256 parallel streams, we deploy two iperf3 processes with 128 parallel streams each using "-P" option. Additionally, we monitor the overall CPU utilization using "mpstat all all" every second throughout the runs, and use a python script to calculate the min, max and average CPU utilization. Figure 6(a) shows the aggregate throughput (sum of the average throughput of each TCP stream) achieved through each interface from a single stream up to 512 parallel streams, and Figure 6(b) shows the overall CPU utilization (percentage of CPU utilization with respect to all cores) observed.

We observe that with increasing number of parallel streams, the throughput achieved through native interface surges from about 90 Mbps up to 3.5 Gbps, and CPU utilization increases to about 23% with peak utilization of about 31%, due to multiple threads deployed per stream. While tunnel interfaces (Geneve, VXLAN and OVS counterparts) exhibit similar throughput compared to native over a single TCP stream, the degradation in throughput widens with multiple streams as we notice the aggregate throughput increase to only about 750 Mbps (1/5x times lesser). However, we observe that CPU utilization consistently increases. The trend in this increased CPU utilization is higher for OVS tunnels, which experience average CPU utilization to go beyond 26 %, and max CPU utilization to 38%. Linux-based Geneve and VXLAN tunnels seem to offer better throughput-to-CPU ratio. The Wireguard encryption interface, which achieved higher throughput with a single stream doesn't keep up the momentum with multiple streams and achieves about 550 Mbps using about 22% of CPU on an average. With IPSec, we observe the least aggregate throughput of about 450 Mbps since it has a bottleneck of utilizing only a single CPU even with parallel streams, and hence observes lesser CPU utilization of about 10%.

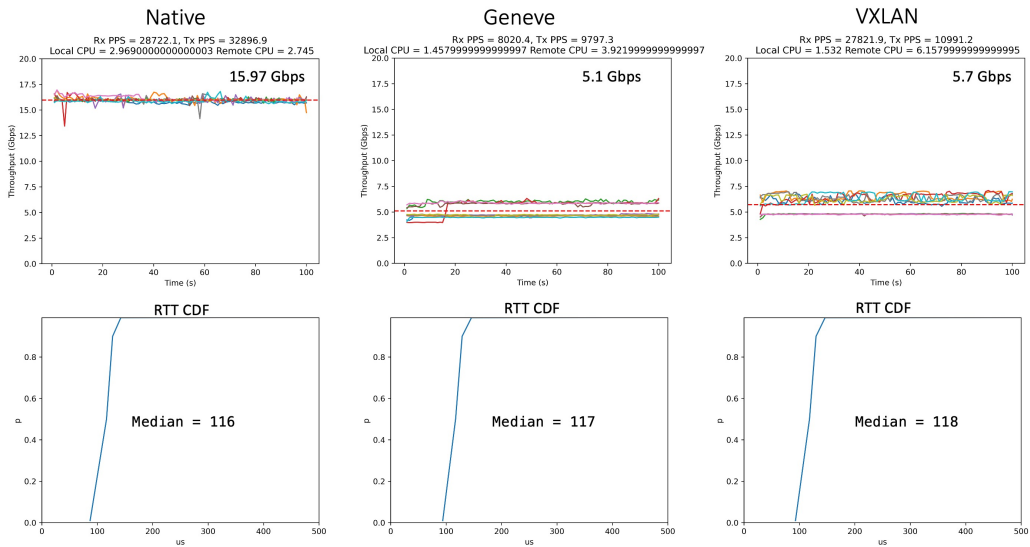

**Fig. 7.** Throughput series and RTT CDF of the Native, Geneve and VXLAN interfaces for a single stream netperf in intra-DC setup

**Take-aways:**

(1) We observe 5x throughput degradation with Geneve, VXLAN and OVS-based tunnels, while observing similar CPU utilization compared to Native interface without tunnels.
(2) OVS-based tunnels consume more CPU cycles while achieving similar throughput compared to linux tunnels.
(3) We observe 7x throughput degradation with encryption using both Wireguard and IPSec.
(4) Wireguard performs marginally better than IPSec due to its higher CPU utilization across multiple cores.

### 3.2 Intra-DC Measurements

We now move on to the measurements on a pair of intra-DC VMs. The VMs are hosted on IBM's Cloud Virtual Servers for VPC (Gen2) VMs[1] (both VMs co-located in Sydney DC). Both VMs are equipped with 16-core Intel Xeon @2.5 GHz. We do not specify any specific amount of network bandwidth to be provisioned, but based on the configuration available, we assume 16Gbps was provisioned.

As performed previously, we again perform 10 separate runs for each interface using netperf (Listing 1 & 2) to observe the throughput and RTT measurements for a single TCP stream. Figure 7, 8 and 9 show the throughput series and the CDF of RTT for a single TCP stream. We observe the native interface provides about 14.5 Gbps with a median RTT of 130 $\mu$s, while Geneve, VXLAN and their OVS counterparts achieve only about a third of throughput as that of the native interface. We attribute this to TCP Segmentation Offload (TSO/GRO) enabled on the VMs. Due to this, the VM's native interface can aggregate TCP packets up to size of 64 K. However, after switching to UDP-based tunnels, TSO/GRO cannot be functional anymore. With native linux tunnels, we observe a very negligible increase in latency. However, with OVS-based tunnels we observe a higher increase in latency of upto 10 us.

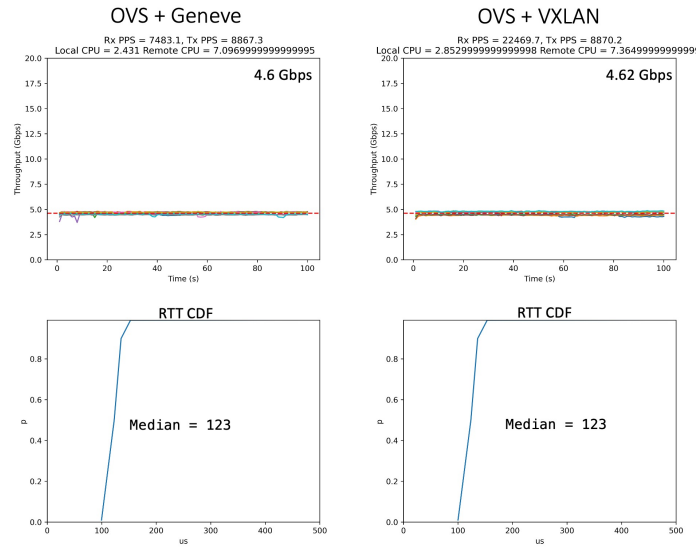

Fig. 8. Throughput series and RTT CDF of the OVS+Geneve and OVS+VXLAN interfaces for a single stream netperf in intra-DC setup

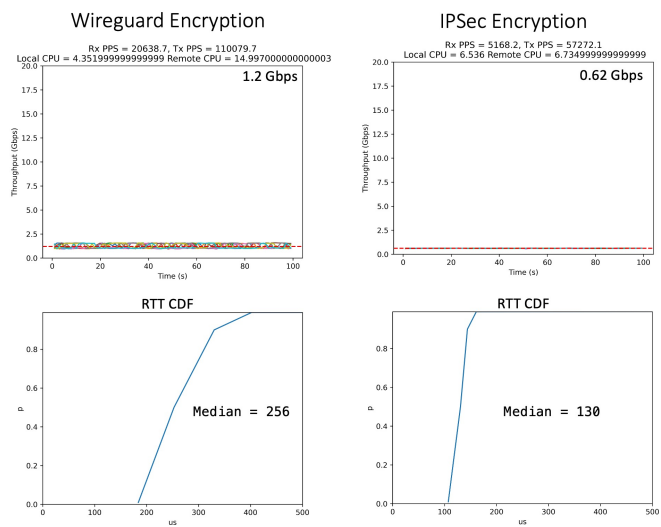

Fig. 9. Throughput series and RTT CDF of the Wireguard and IPSec interfaces for a single stream netperf in intra-DC setup

With encryption, we observe a drastic drop in throughput as seen in Figure 9. While Wireguard interface achieves 1.2 Gbps on an average, IPSec achieves only about 620 Mbps. The median RTT doubles with Wireguard interface and increases by about 10s of us with IPSec.

Next, to understand if we can mitigate the impact of TSO/GRO to achieve parity in performance with tunnels, we increase the MTU (Maximum Transmission Unit) of the Native interface to 9000 bytes, and adjust the MTU of the tunnel/encryption interfaces accordingly after adjusting for the additional header size (8940 for tunnel interfaces and 8920 for Wireguard interface). Note that the IPSec's ESP packets were getting dropped due to VM restrictions, and hence we used IPSec with UDP encapsulation. We show the results in Figure 10, 11 and 12. We observe a slight increase in average throughput of native interface by increasing the MTU to 9000 bytes. Interestingly, Geneve and its OVS counterpart achieve parity in performance by increasing packet size to 8940 bytes. This highlights that tunnelling will lose the advantage of segmentation offload (TSO/GRO) unless MTU is increased accordingly. While Wireguard's performance rises by about 4x times, IPSec observes only 1.5x times improvement in performance.

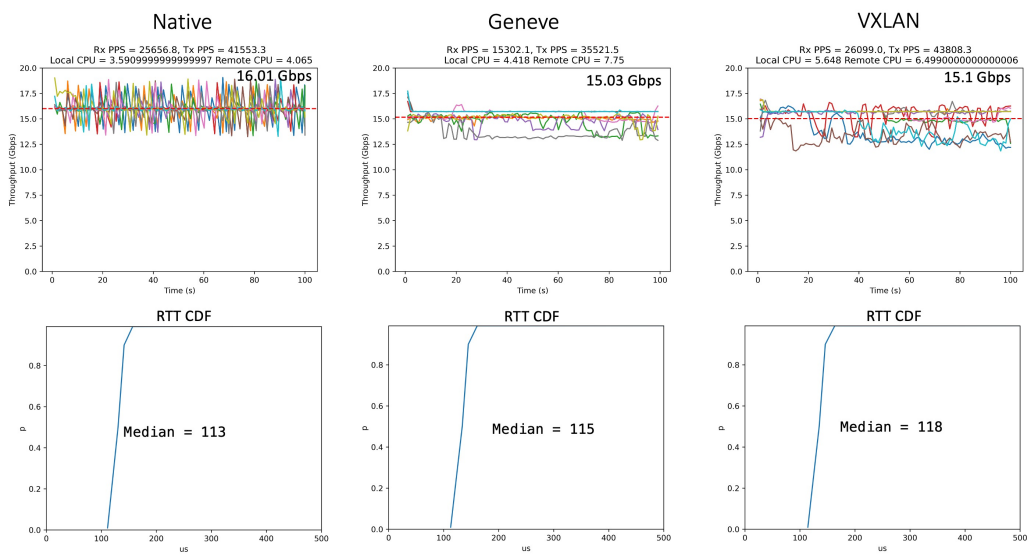

Fig. 10. Throughput series and RTT CDF of the Native, Geneve and VXLAN interfaces for a single stream netperf with increased MTU of 9000 bytes in intra-DC setup

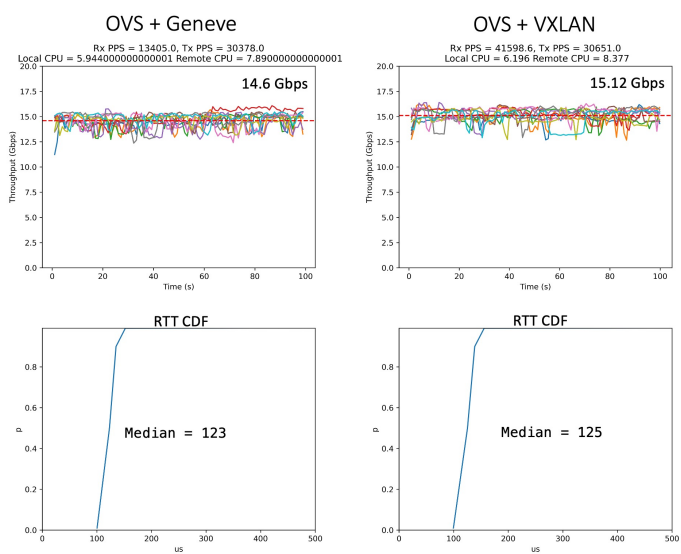

Fig. 11. Throughput series and RTT CDF of the OVS+Geneve and OVS+VXLAN interfaces for a single stream netperf with increased MTU of 9000 bytes in intra-DC setup

To understand this performance improvements, we plot the average CPU utilization observed by netperf in Figure 13. We observe a marginal increase in CPU utilization with tunnel interfaces, and a significant increase in CPU utilization with Wireguard. IPSec observes only a marginal increase in CPU utilization and hence it doesn't achieve higher throughput.

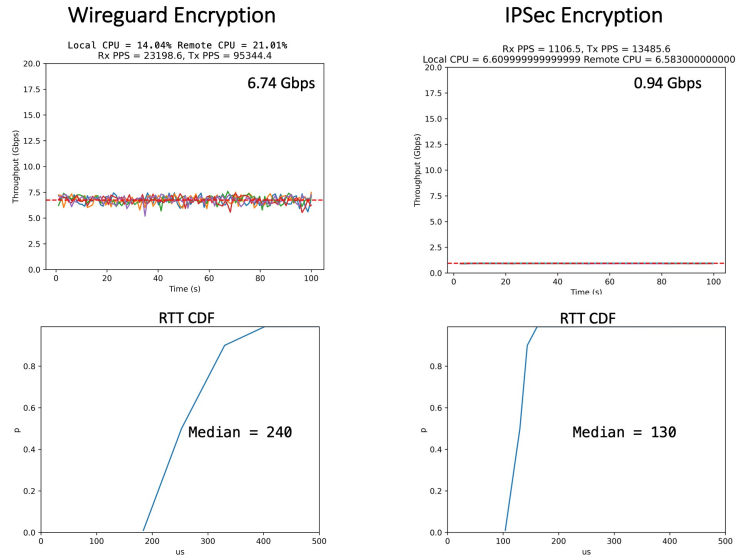

Fig. 12. Throughput series and RTT CDF of the Wireguard and IPSec interfaces for a single stream netperf with increased MTU of 9000 bytes in intra-DC setup

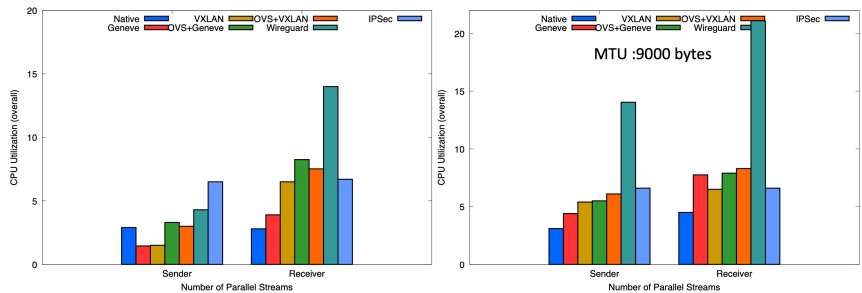

Fig. 13. Average CPU utilization observed by netperf sender/receiver during the measurements in intra-DC setup

Next, we scale the measurement beyond a single TCP Stream to multiple parallel streams (1500-byte MTU) using iperf3. We report the throughput observed by using parallel streams upto 512 in Figure 14(a), and the corresponding CPU utilization incurred in Figure 14(b). We observe that the native interface reaches line-rate and peak at close to 16 Gbps. We also observe that with increase in number of parallel streams beyond 64 flows, we observe a drastic drop in throughput. This could be attributed to the contention among the various parallel streams. This behaviour is consistent with all tunnel interfaces, while encryption interfaces do not see this effect due to pre-existing high processing of encryption. Another observation is VXLAN interface performs much better (almost 2x times) than Geneve interface, while it incurs lesser CPU than Geneve. While they achieve similar throughput with a single stream, VXLAN performs much better when increasing the number of parallel streams. One possibility is that of some specific VXLAN offload feature provided by the cloud. Other possibility is that VXLAN kernel module is highly optimized. However, we do not observe the same trend with OVS+VXLAN, which observes higher CPU utilization and slightly lesser throughput than native VXLAN interface.

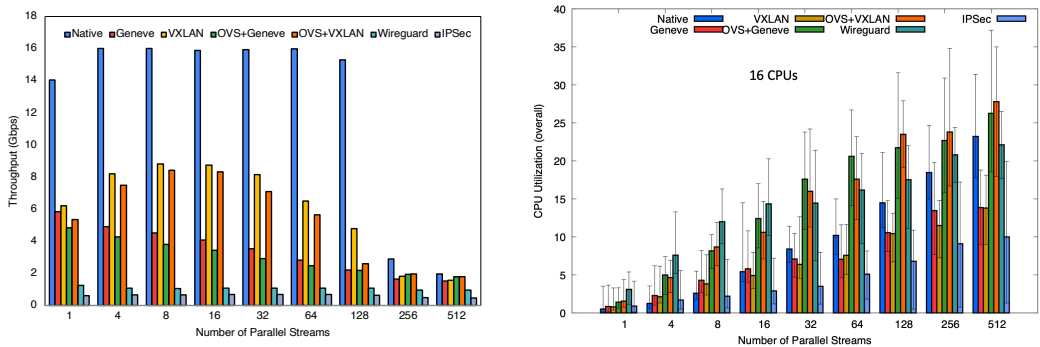

(a) Aggregate throughput observed using iperf3 for multiple parallel TCP streams

(b) Average CPU utilization observed using Mpstat for multiple parallel TCP streams

Fig. 14. Throughput and CPU utilization for multiple parallel TCP streams

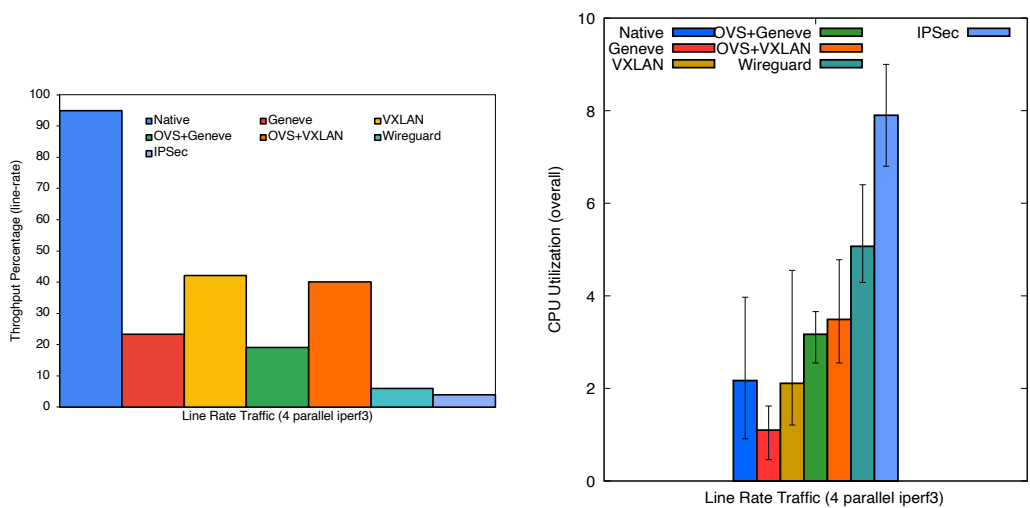

(a) Aggregate throughput observed using iperf3 for 4 parallel streams

(b) Average CPU utilization observed using Mpstat for 4 parallel TCP streams

Fig. 15. Throughput and CPU utilization for four parallel TCP streams at line-rate

Next to avoid the problem of contention of too many flows, we run 4 parallel iperf3 without any parallel streams (no -P option). We chose 4 iperf3s by incrementally increasing the number of iperf3 processes to achieve line-rate peak before diminishing. We plot the percentage of line-rate throughput observed by each interfaces in Figure 15(a) and the corresponding CPU utilization in Figure 15(b). While we observe 40% of line-rate traffic with VXLAN, only 20% could be achieved using GENEVE. Wireguard and IPSec achieve less than 10% of line-rate, while consuming significantly higher amount of CPU.

**Take-aways:**

(1) Throughput with Geneve/VXLAN tunnels reduces by about 3x times with normal MTU of 1500 byte packets, because native traffic leverages TSO/GRO features offered by the next-gen VMs with offload features.

(2) The throughput decrease can be compensated by enabling jumbo frames (9000-byte packets).

(3) While Wireguard's throughput increases by 4x with jumbo frames, IPSec's performance increases very marginally (≈1.5x).

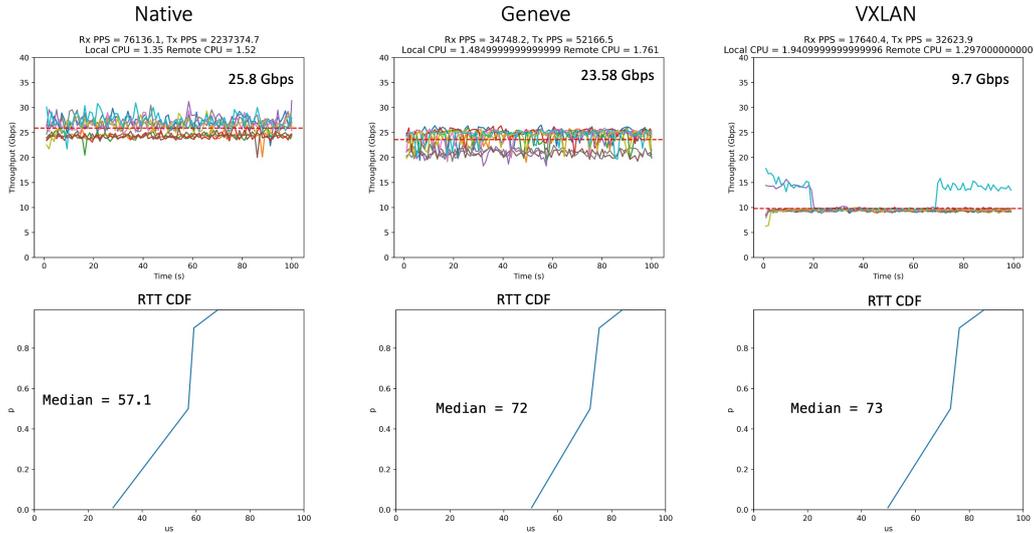

Fig. 16. Throughput series and RTT CDF of the Native, Geneve and VXLAN interfaces for a single stream netperf in intra-rack setup

(4) With increasing number of parallel streams, native VXLAN tunnels performs much better offering more throughput for lesser CPU utilization.

(5) The same trend is not observed by OVS-based VXLAN tunnel which consumes double the CPU utilization.

(6) At optimum traffic (4 parallel iperf3s), Geneve tunnels achieve only 23% of line-rate, while VXLAN tunnels achieve 41% of line-rate.

(7) At optimum traffic (4 parallel iperf3s), Wireguard and IPSec encryption achieve only 6% and 4% of line-rate respectively.

### 3.3 Intra-rack Measurements

We start the measurements on a pair of intra-rack bare-metal servers. The bare-metals are equipped with Both VMs are equipped with 80-core Intel Xeon @2.5 GHz connected using a 40G DAC cable via Mellanox Connect-X 5 NICs [7].

Similar to previous measurements, we again perform 10 separate runs for each interface using netperf for single TCP stream. Figure 16, 17 and 18 show the throughput series and the CDF of RTT achieved throughout the netperf measurements. We observe the native interface achieves about 26Gbps with a single stream of Netperf. Interestingly, we observe about 23.5 Gbps with Geneve. This is because the Mellanox ConnectX-5 NIC support UDP tunnel offload. Hence, the Geneve UDP packets being sent off the interface were 64K in size, and the segmentation was happening in the NIC. However, we did not observe native VXLAN tunnels offer the same benefit, since we did not observe the VXLAN tunnel packet segmentation being offloaded to the NIC as the packets were being sent as 1500-byte packets[2]. However, we observe that OVS-based tunnels were being offloaded to the NIC for segmentation. Thus, they observe better performance than native VXLAN tunnels without offload.

The intra-rack bare-metal setup gives the best way to observe latency overhead due to tunnel and encryption since the network latency is fixed without queuing at the switches. We observe latency increase of about 15 microseconds with Geneve and VXLAN tunnels, and an increase of 25 microseconds with OVS counterparts.

---

[2]We verified using TCPdump packet capture. Do note that we did not change any settings explicitly in the NIC, nor did we add/change the driver of the Mellanox NIC.

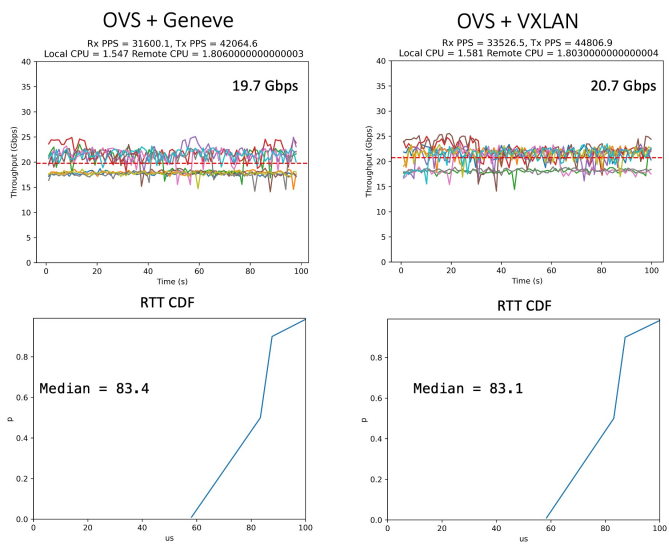

Fig. 17. Throughput series and RTT CDF of the OVS+Geneve and OVS+VXLAN interfaces for a single stream netperf in intra-rack setup

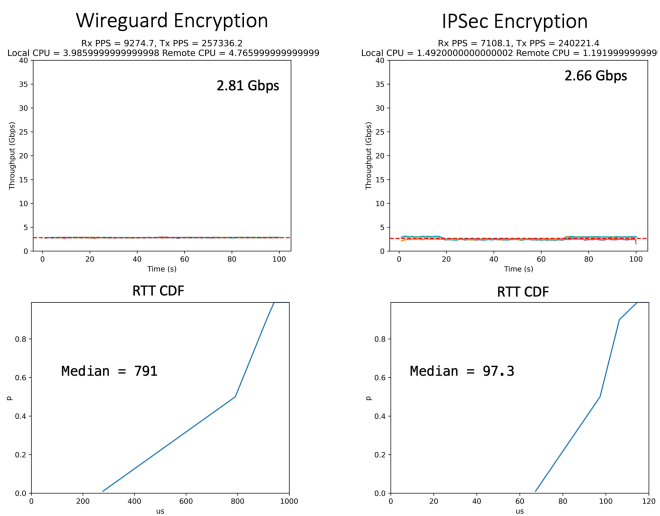

Fig. 18. Throughput series and RTT CDF of the Wireguard and IPSec interfaces for a single stream netperf in intra-rack setup

Even with bare-metal servers, we observe encryption sees 10x drop in performance. While Wireguard's latency increases by 100 times, IPSec latency increases by only 40 microseconds. However, since IPSec works on a single core, it doesn't utilize CPU optimally. Next, we perform the same set of experiments after changing the MTU of the network interface to 9000 bytes (jumbo frames) as shown in Figure 19, 20 and 21. We observe that subsequently, the throughput of a single flow increases significantly. VXLAN interface still under performs compared to Geneve native interface. The OVS counterparts perform consistently.

We plot the CPU utilization observed by netperf in Figure 22. We observe that tunnelling on a single stream of netperf does not increase CPU utilization. Wireguard as observed previously observes the highest CPU utilization for both normal frames and jumbo frames.

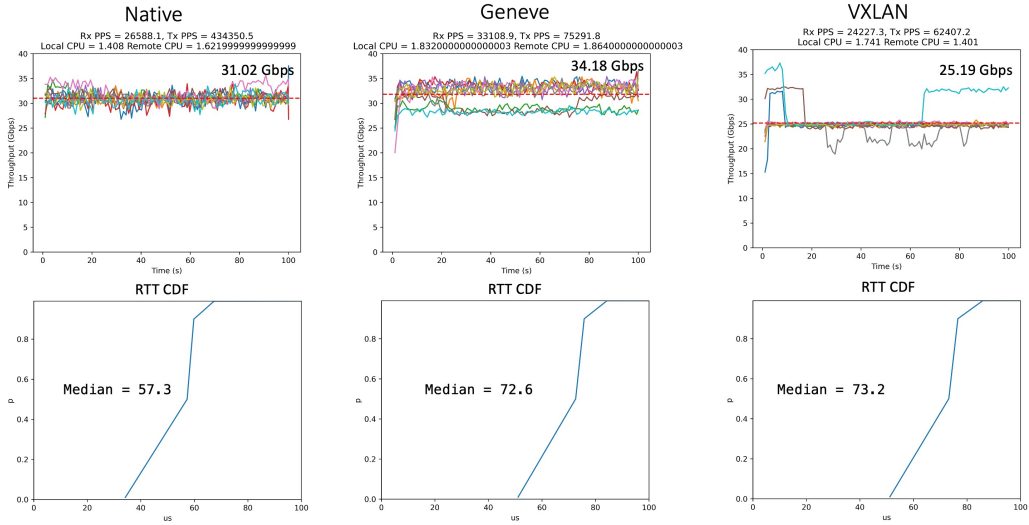

Fig. 19. Throughput series and RTT CDF of the Native, Geneve and VXLAN interfaces for a single stream netperf in intra-rack setup

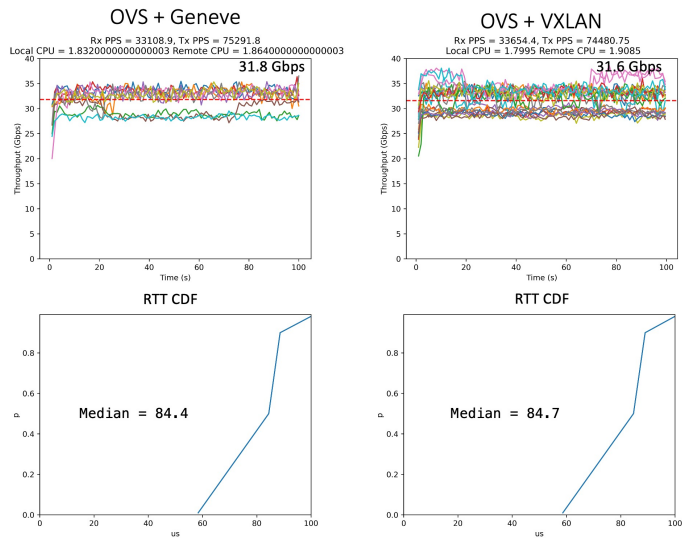

Fig. 20. Throughput series and RTT CDF of the OVS+Geneve and OVS+VXLAN interfaces for a single stream netperf in intra-rack setup

Next, we scale the measurement beyond a single TCP Stream to multiple parallel streams (1500-byte MTU) using iperf3. We report the throughput observed by using parallel streams upto 512 in Figure 23(a), and the corresponding CPU utilization incurred in Figure 23(b). Our main observation is that beyond 256 flows, since the number of cores are higher, throughput increases significantly for tunnel traffic. However, the encryption interfaces do not see a significant increase. However, what we notice is that Wireguard interface throughput starts to decrease with increasing number of parallel flows, while we notice the opposite with IPSec. This could be explained by IPSec 's operation is limited to a single CPU, and in the bare-metal scenario, a single CPU performance is good. However, Wireguard's threads could not leverage multiple CPUs effectively due to potentially shared states and increased contentions [17]. Figure 23(b) points to decreasing CPU utilization with Wireguard pointing to more time spent in possibly waiting for locks on shared states, While

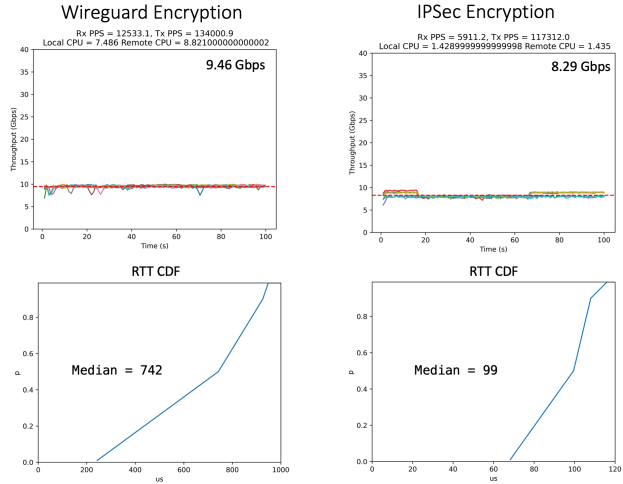

Fig. 21. Throughput series and RTT CDF of the Wireguard and IPSec interfaces for a single stream netperf in intra-rack setup

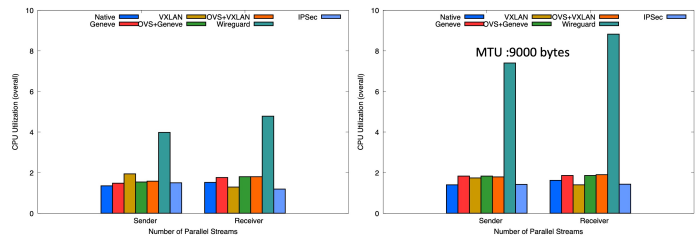

Fig. 22. Throughput and CPU utilization for multiple parallel TCP streams

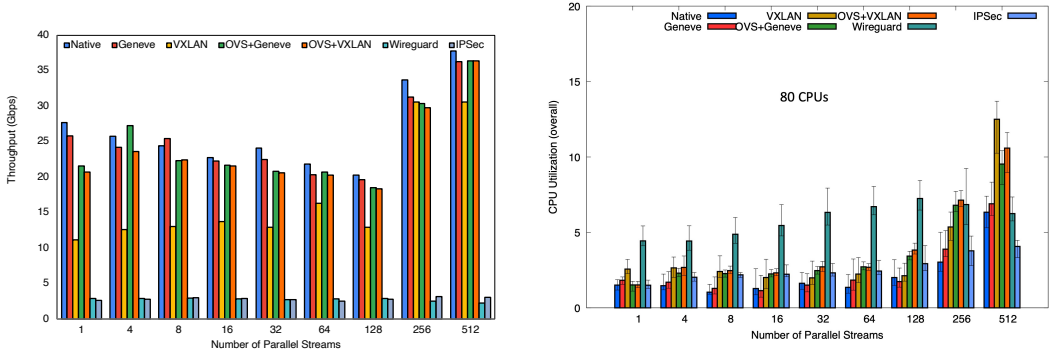

(a) Aggregate throughput observed using iperf3 for multiple parallel TCP streams

(b) Average CPU utilization observed using Mpstat for multiple parallel TCP streams

Fig. 23. Throughput and CPU utilization for multiple parallel TCP streams

IPSec CPU utilization increased. We consistently observe that increase in CPU utilization due to native Geneve tunnels is very minimal. This is primarily due to segmentation offload to NICs. On the other hand, VXLAN's CPU utilization increases significantly with increase in flows.

Next, we run 4 parallel iperf3 without any parallel streams (no -P option). We chose 4 iperf3s as previously by incrementally increasing the number of iperf3 processes to achieve line-rate peak before diminishing. We plot the percentage of line-rate throughput observed by each interface in Figure 24(a) and the corresponding CPU utilization in Figure 24(b). We observe that all tunnel

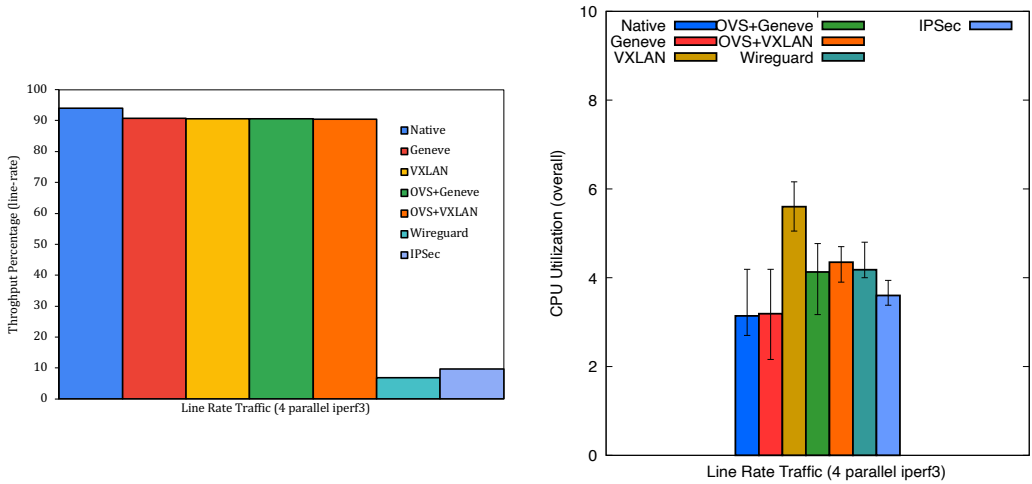

(a) Aggregate throughput observed using iperf3 for four parallel TCP streams

(b) Average CPU utilization observed using Mpstat for four parallel TCP streams

Fig. 24. Throughput and CPU utilization for four parallel TCP streams at line-rate

interfaces achieve close to line-rate performance. The small depreciation in throughput is because of extra-header overhead incurred in the tunnel packets. Although, VXLAN tunnel did not get offloaded, it achieves line-rate with higher CPU utilization compared to others. Wireguard achieves 6.8% of line-rate and IPSec achieves 9.6% of line-rate. The reason for Wireguard's lower performance could be because of contention among shared states as observed in lower CPU utilization of Wireguard in Figure 24(b). Interestingly, VXLAN tunneling incurs the most CPU since it couldn't be offloaded to NIC.

**Take-aways:**

(1) Native Geneve tunnels get accelerated with recent NICs like Mellanox ConnectX-5 due to UDP tunnel segmentation.

(2) VXLAN tunnels however did not get accelerated by UDP tunnel segmentation offload feature in NIC.

(3) OVS-based tunnels get offloaded to the NIC for segmentation.

(4) IPSec performs better than Wireguard if single-core CPU performance is good even without IPSec offload.

## 4 CONCLUSION & FUTURE WORK

We conducted measurements in three distinctly different setups and observed varied performance of different tunnel encapsulations and encryption methods. In VM-based setup, we observed VXLAN tunnel interfaces consistently outperform Geneve. However, when the NIC supports UDP tunnel segmentation native Geneve interfaces performed better with fewer CPU cycles. One common observation is that OVS-based tunnels generally consume more CPU cycles while delivering lesser throughput compared to native Linux tunnels. When it comes to encryption, we observe Wireguard, in general, performs better than IPSec, since it runs on multiple CPU threads. However, IPSec performs better when the number of parallel flows increases, which increases the contention of shared resources (leading to bottleneck at Wireguard).

This study highlights that we cannot take a *"one-size-fits-all"* approach to choose which of the tunnel/encryption methodologies to be used to connect sites and will need to be decided based on the deployment setting and the use-case. It also highlights which deployment setting is most suitable for a particular tunnel/encryption methodology.

**Management**. Native linux-based tunnels rely on separate interfaces on either side being created for an end-to-end tunnel. While this is not a problem in a normal setting, in the context of Multi Cloud Networking, such tunnels need to scale and connect to more clusters and denser workloads. Clearly, linux-based tunnels will lead to an explosion of interfaces being created and will further lead to a tunnel management nightmare for network operators. Hence, there is a need to have a formal tunnel management system that can provide rule-based protocol-agnostic end-to-end tunnel management. To this end, the emergence of eBPF(Extended Berkeley Packet Filter) provides a promising opportunity in developing a rule-based tunnel-management system in-kernel by offloading rule-based lookup, header manipulations to the faster XDP-based datapath [16]. Additionally, as NICs supporting XDP offload continue to expand in the market, it creates a novel opportunity for developing systems that operate at line-rate without significant performance downgrades.

**Future Work**. In the future, we plan to develop a single end-to-end tunnel using eBPF programs that could be attached at tc/XDP hook points and compare its performance with the current set of benchmarks. We also plan our to include other tunnel methodologies such as L2TP, SSH, GRE tunnels and IPSec's transport mode to our study. Additionally, we plan to broaden the study of the performance of tunnels with more realistic traffic based on micro-services, which usually have a skewed distribution of packet size, with the majority of packets <200 bytes [18]. Of course, such a scenario presents an even more challenging situation where the tunnel header overhead of 50-60 bytes inflates each packet size by about 30-40%. We plan to explore if multiple packets destined to the same cluster could be combined with a single tunnel header.

## REFERENCES

[1] A New Generation of IBM Cloud Virtual Servers for VPC. https://www.ibm.com/cloud/blog/announcements/next-gen-virtual-servers-vpc.
[2] Aviatrix Site2Site IPSec. https://docs.aviatrix.io/HowTos/site2cloud.html.
[3] Cloud Strategy, Gartner IT ICOS Conference 2020. https://www.gartner.com/en/conferences/apac/infrastructure-operations-cloud-india/featured-topics/cloud.
[4] iperf. https://iperf.fr/.
[5] Libreswan VPN software. https://libreswan.org/.
[6] Meeting Data Compliance with a Wave of New Privacy Regulations: GDPR, CCPA, PIPEDA, POPI, LGPD, HIPAA, PCI-DSS, and More. https://cloud.netapp.com/blog/data-compliance-regulations-hipaa-gdpr-and-pci-dss.
[7] Mellanox ConnectX-5. https://www.nvidia.com/en-in/networking/ethernet/connectx-5/.
[8] mpstat. https://man7.org/linux/man-pages/man1/mpstat.1.html.
[9] Netlink - netlink library for go. https://www.github.com/vishvananda/netlink.
[10] Netperf. https://github.com/HewlettPackard/netperf.
[11] Open VSwitch. openvswitch.org.
[12] OVN – Geneve vs VXLAN, Does it Matter? https://blog.russellbryant.net/2017/05/30/ovn-geneve-vs-vxlan-does-it-matter/.
[13] Softlayer's new name : IBM Cloud. https://www.ibm.com/in-en/cloud/info/softlayer-is-now-ibm-cloud.
[14] Submariner Gateway Engine. https://submariner.io/getting-started/architecture/gateway-engine/.
[15] J. A. Donenfeld. Wireguard: Next generation kernel network tunnel. https://www.wireguard.com/papers/wireguard.pdf.
[16] T. Høiland-Jørgensen, J. D. Brouer, D. Borkmann, J. Fastabend, T. Herbert, D. Ahern, and D. Miller. The express data path: Fast programmable packet processing in the operating system kernel. In *CoNEXT*, 2018.
[17] M. Pudelko, P. Emmerich, S. Gallenmüller, and G. Carle. Performance analysis of vpn gateways. In *IFIP Networking Conference (Networking)*, pages 325–333, 2020.
[18] Q. Zhang, V. Liu, H. Zeng, and A. Krishnamurthy. High-resolution measurement of data center microbursts. In *IMC*, 2017.